\documentclass{article}
\usepackage{fleqn}
\usepackage{epsfig}
\usepackage{amsmath,amssymb}
\begin{document}

\begin{center}
{\bf\Large Statistical system with fantom scalar interaction. II. Macroscopic Equations and Cosmological Models.}\\[12pt]
Yu.G. Ignatyev, A.A. Agathonov and D.Yu. Ignatyev\\
Kazan Federal University,\\ Kremlyovskaya str., 35,
Kazan 420008, Russia
\end{center}

{\bf keywords}: Early Universe, fantom scalar interaction, relativistic kinetics,
cosmological acceleration, numerical simulation.

\begin{abstract}
Based on the proposed earlier by the Author approach to macroscopic description of scalar interaction, this paper develops the macroscopic model of relativistic plasma with a fantom scalar interaction of elementary particles. In the article the macroscopic equations for a statistical system with scalar interaction of particles are obtained and the complete set of macroscopic equations describing cosmological models is built.
\end{abstract}

\section{Kinetic And Transport Equations}
In the previous article \cite{Ignat14} there were stated and investigated the canonical and the Lagrange common-relativistic
equations of motion for a scalar charged particle in scalar field without any restriction on the sign of particle effective mass.
As was seen, the cited canonical equations are stated unambiguously accurate to non-existing isomorphisms at choice of a scalar Hamilton function.
 The structure of the relativistic kinetic theory  (see e.g. \cite{Yubook1}), as was noticed earlier in \cite{Ignat14}, allows the construction of kinetic equations at known canonical equations and also it allows us to unambiguously construct the strict macroscopic consequences of the kinetic equations (so-called {\it transport equations}) at known Lagrange function; these consequences again un\-ambi\-gu\-ous\-ly lead to the self-consistent set of Einstein equations, equation of the scalar field with a source and ma\-cro\-sco\-pic conservation laws for the statistical system of charged particles. Such equations in dif\-ferent representations were constructed by one of the Authors first in 1983 \cite{Ignatev2,Ignatev3,Ignatev4} for classical scalar fields \cite{YuNewScalar1,YuNewScalar2,YuNewScalar3}. Let us note that obtained results in all that works were applied to the case of positively defined mass of particles. As is shown in \cite{Ignat14}, the microscopic dynamics of particles in scalar field does not impose any requirements on the sign of the effective mass of particles. Nevertheless let us write out basic relations of the kinetic model with an account of this generalization and highlight those elements of the kinetic theory's construction in which its invariance with respect to the change of a mass function's sign is broken.

\subsection{The Distribution Function And Its Moments}
Let $F(x,P)$ is an invariant function of particles distribution in 8-dimensional
phase space and let $\psi (x,P)$ is a certain tensor function of dy\-namic variables $(x,P)$.
According to \cite{Yubook1} each tensor dy\-namic function can be associated with a macroscopic flux density:
\begin{eqnarray} \label{GrindEQ__35_} \Psi ^{i} (x)&=&
\int\limits_{P(x)} F(x,P)\psi (x,P)\frac{\partial H}{\partial P_{i} } dP\nonumber\\
&\equiv & m_{*}^{-1} \int\limits_{P(x)} F(x,P)\psi (x,P)P^{i} dP, \end{eqnarray}
Let us define according to \eqref{GrindEQ__35_} moments with respect to the distribution $F(x,P)$ \cite{Yubook1}:
\begin{equation} \label{GrindEQ__36_}
n^{i} (x)=\!\!\!\int\limits_{P(x)}\!\!\! F(x,P)\frac{\partial H}{\partial P_{i} } dP
\equiv\!\! m_{*}^{-1} \int\limits_{P(x)}\!\!\! F(x,P)P^{i} dP,
\end{equation}
vector of particle number's flux density \footnote{Number vector according to J.L. Synge \cite{Sing}.}, so that:
\begin{equation} \label{GrindEQ__37_} n^{i} =nv^{i} , \end{equation}
where $v^{i} $ is a timelike unit vector of kinematic macroscopic velocity of particles:
\begin{equation} \label{GrindEQ__38_} n=\sqrt{(n,n)} . \end{equation}
Next:
\begin{eqnarray} \label{GrindEQ__39_} T_{p}^{ik} (x)&=&\int\limits_{P(x)} F(x,P)P^{i} %
\frac{\partial H}{\partial P_{k} } dP\nonumber\\
&\equiv & m_{*}^{-1} \int\limits_{P(x)} F(x,P)P^{i} P^{k} dP,
\end{eqnarray}
is a macroscopic energy-momentum tensor (EMT). Trace of this tensor can be calculated by means of normalization ratio \cite{Ignat14}
\begin{eqnarray}\label{GrindEQ__8_}
H(x,P)=\frac{1}{2} \left[m_*^{-1}(x)(P,P)-m_*(x) \right]=0\\
\Rightarrow(P,P)=m_*^2\Rightarrow
\end{eqnarray}
\begin{equation} \label{GrindEQ__40_} T_{p} \equiv g_{ik} T_{p}^{ik} = %
m_{*} \int\limits _{P(x)} F(x,P)dP. \end{equation}
Next, an invariant volume element of 4-dimensional momentum space in expressions %
\eqref{GrindEQ__36_}, \eqref{GrindEQ__39_} in chosen system of units is \cite{Yubook1}:
\begin{equation} \label{GrindEQ__41_} dP=\frac{2S+1}{(2\pi )^{3} \sqrt{-g} } dP_{1} \wedge dP_{2} \wedge dP_{3} \wedge dP_{4} , \end{equation}
where $S$ is a spin of particle. The sole way to connect  the invariant 8-dimensional distribution function
$F(x,P)$ to the 7-dimensional one, $f(x,P)$, which is defined on a mass surface, could be realized only by means of
function of Hamiltonian since Hamilton function is conserved at movement along
the particle's world line \cite{Ignat14}, \cite{YuStatistic}. %
Thus an invariant 8-dimensional distribution function $F(x,P)$, singular on a mass surface \eqref{GrindEQ__8_}
is connected to a nonsingular 7-dimensional distribution function $f(x,P)$ through $\delta $-function by means of the relation \cite{Yubook1}:
\begin{eqnarray} \label{GrindEQ__42_} F(x,P)=&f(x,P)\delta (H)=\nonumber\\%
 & |m_{*}| \frac{\delta (P_{4} -P_{4}^{+} )}{P_{+}^{4} }f(x,P) , \end{eqnarray}
where $P_{4}^{+} $ is a positive root of the normalization equation \eqref{GrindEQ__8_};
 $P_{+}^{4} =g^{4k} P_{k}^{+} $ is a value of contravariant
 momentum component corresponding to that root. In local Lorentz frame of reference it is:
\begin{equation} \label{GrindEQ__43_} P_{+}^{4} =\sqrt{m_{*}^{2} +P^{2} } , \end{equation}
where $P^{2} =\sum _{\alpha =1}^{3} (P^{\alpha } )^{2} $ is a quadrate of physical momentum.
Thus we obtain an invariant volume element of 3-dimensional momentum space:
\begin{eqnarray} \label{GrindEQ__44_} dP_{+} =   |m_{*}| \frac{2S+1}{(2\pi )^{3}
\sqrt{-g} } \frac{dP_{1} \wedge dP_{2} \wedge dP_{3} }{P_{+}^{4} }\nonumber \\   \equiv |m_{*}| dP_{0} ,
\end{eqnarray}
where
\begin{eqnarray} \label{GrindEQ__45_} dP_{0} =\sqrt{-g} \frac{2S+1}{(2\pi )^{3} } \frac{dP^{1} dP^{2} dP^{3} }{P_{4}^{+} } \\ \equiv \sqrt{-g} \frac{2S+1}{(2\pi )^{3} } \frac{d^{3} P}{P_{4}^{+} } . \end{eqnarray}
 Then \eqref{GrindEQ__36_}, \eqref{GrindEQ__39_} and \eqref{GrindEQ__40_} take the following form
  (for simplicity we omit a summation by particle sorts):
\begin{eqnarray} \label{GrindEQ__46_} n^{i} (x)=&{\rm sgn}(m_*)\times\hskip 3.1cm \nonumber \\
&\frac{2S+1}{(2\pi )^{3} } \int\limits _{P(x)} f(x,P)P^{i} \sqrt{-g} \; \frac{d^{3} P}{P_{4}^{+} } ;\\[0pt]
\label{GrindEQ__47_} T_{p}^{ik} (x)=&{\rm sgn}(m_*)\times\hskip 2.9cm \nonumber \\
&\frac{2S+1}{(2\pi )^{3} } \int\limits _{P(x)} f(x,P)P^{i} P^{k} \sqrt{-g} \; \frac{d^{3} P}{P_{4}^{+} } ;\\[0pt]
\label{GrindEQ__48_} T_{p} =&{\rm sgn}(m_*)\times\hskip 3.1cm  \nonumber \\
&\frac{2S+1}{(2\pi )^{3} } m_{*}^{2} \int\limits _{P(x)} f(x,P)\sqrt{-g} \frac{d^{3} P}{P_{4}^{+} } . \end{eqnarray}
Thus macroscopic fluxes are explicitly defined by the sign of an effective mass $m_{*}$, %
specifically particle energy density becomes negative once
the sign of a mass function is changed, which case we now can't allow due to conservatism\footnote{Let us note that
value of an effective mass modulus $|m_*|$ appeared determine as a result of resolution of $\delta$ - %
Dirac function of invariant Hamiltonian with respect to the momentums on mass surface.}. %
Thus further {\it as a consequence of the requirements of the standard statistical theory
} we choose a constant-sign function of an effective mass:
 \begin{equation}\label{|m|}
 m_*=|m+q\Phi|.
 \end{equation}
 Therefore further we omit factor ${\rm sgn}(m_*)$ in (\ref{GrindEQ__46_}) -- (\ref{GrindEQ__48_}).

\subsection{Kinetic And Transport Equations}
As a result of local correspondence principle and suggestion of particle collision being 4-pointed one, in each act of interparticle interaction is conserved a generalized momentum of the interacting particles system:
\begin{equation} \label{GrindEQ__49_} \sum _{I} P_{i} =\sum _{F} P'_{i} , \end{equation}
where the summation is carried our by all initial, $P_{i} $, and final, $P'_{i} $, %
conditions. Let the following reactions proceed in plasma:
\begin{equation} \label{GrindEQ__50_} \sum _{A=1}^{m} \nu _{A} a_{A} {\rm \rightleftarrows }\sum _{B=1}^{m'} \nu '_{B} a'_{B} , \end{equation}
where $a_{A} $ are particles characters and $\nu _{A} $ are their numbers. Thus generalized momentums
of the initial and final conditions are equal to:
\begin{equation} \label{GrindEQ__51_}
P_I=\sum\limits_{A=1}^{m}\sum\limits_{\alpha}^{\nu_A}P^\alpha_A=P_F.
\end{equation}
Particles' distribution functions are determined by the invariant kinetic equations \cite{Ignatev3}:
\begin{equation} \label{GrindEQ__52_} [H_{a} ,f_{a} ]=I_{a} (x,P_{a} ), \end{equation}
where $J_{a} (x,P_{a} )$ is an integral of collisions:
\begin{eqnarray} \label{GrindEQ__53_} I_{a} (x,P_{a} )=-\sum  \nu _{a} \int  '_{a} \delta ^{4} (P_{F} -P_{I} )\times \nonumber \\
W_{IF} (Z_{IF} -Z_{FI} )\prod _{I,F} 'dP; \end{eqnarray}
\[W_{FI} =(2\pi )^{4} |M_{IF} |^{2} 2^{-\sum  \nu _{A} +\sum  \nu '_{b} }\]
is a scattering matrix of channel of reactions \eqref{GrindEQ__50_} ($|M_{IF} |$ are invariant scattering amplitudes);
\begin{eqnarray}
Z_{IF} =\prod _{I} f(P_{A}^{\alpha } )\prod _{F} [1\pm f(P_{B}^{\alpha '} )];\nonumber\\
Z_{FI} =\prod _{I} [1\pm f(P_{A}^{\alpha } )]\prod _{F} f(P_{B}^{\alpha '} ),\nonumber
\end{eqnarray}
sign ``+'' corresponds to bosons and ``-'' corresponds to fermions (see details in %
\cite{Yubook1,Ignatev3,Ignatev4}). {\it Strict con\-se\-qu\-ences} of general-relativistic kinetic equations \eqref{GrindEQ__52_}
are transport equations of dynamic variables $\Psi _{a} (x,P_{a} )$ \cite{Yubook1}:
\begin{equation}\label{GrindEQ__54_}{\displaystyle \begin{array}{l}
\nabla _{i} \sum _{a}\int\limits _{P(x)}\Psi _{a} F_{a}
\frac{\partial H_{a} }{\partial P_{i} } dP_{a} -\\

\sum _{a} \int\limits _{P(x)} F_{a} [H_{a} ,\Psi _{a} ]dP_{a} =\\
-\sum\limits _{by\; chanels} \int  \left(\sum _{A=1}^{m} \nu _{A} \Psi _{A} -
\sum _{B=1}^{m'} \nu '_{B} \Psi '_{B} \right)\\
\delta ^{4} (P_{F} -P_{I} )(Z_{IF} W_{IF} -Z_{FI} W_{FI} )\prod _{I,F} dP,
\end{array}}
\end{equation}
where $dP_\Sigma=\prod _{I,F} dP$ summation is carried out by all channels of reactions \eqref{GrindEQ__50_}.
Supposing $\Psi _{a} =P^{k} $ in \eqref{GrindEQ__54_} with an account of (13)\cite{Ignat14}, \eqref{GrindEQ__49_} è \eqref{GrindEQ__51_}
we obtain the plasma's energy-momentum transport equation:
\begin{equation} \label{GrindEQ__57_} \nabla _{k} T_{p}^{ik} -\sigma \nabla ^{i} \Phi =0, \end{equation}
where a  \textit{ scalar charge density}, $\sigma $, is introduced [5]:
\begin{equation} \label{GrindEQ__58_} \sigma =\sum _{a} \frac{2S+1}{(2\pi )^{3} } q(m+q\Phi) \int\limits _{P(x)} f(x,P)\sqrt{-g} \; \frac{d^{3} P}{P_{4}^{+} } , \end{equation}

It should be noted that form of (EMT) \eqref{GrindEQ__39_} or \eqref{GrindEQ__47_}, as well as the one of scalar charge density \eqref{GrindEQ__58_} obtained for scalar charged particles in [4], at defined Hamilton function is an unambiguous consequence of the suggestion about total momentum conservation in local particle collisions.

\subsection{A Complete Set Of Macroscopic Equations}
The complete set of macroscopic equations consists first of all from Einstein equations:
\begin{equation}\label{Einst_Scalar}
R^{ik}-\frac{1}{2}Rg^{ik}=8\pi (T^{ik}_p+T^{ik}_s),
\end{equation}
where $T^{ik}_p$ is a determined earlier EMT of the statistical system and $T^{ik}_s$ is an EMT of a scalar field:
\begin{equation}\label{Tik_s}
T_{s}^{ik} =\frac{\epsilon_1}{8\pi } \left[2\Phi ^{,i} \Phi ^{,k} -g^{ik} \Phi _{,j} \Phi ^{,j} +\epsilon_2 m_{s}^{2} g^{ik} \Phi ^{2} \right],
\end{equation}
where for a classic scalar field $\epsilon_2=1$, for fantom scalar field %
$\epsilon_2=-1$; for a field with a repulsion of like charge particles  $\epsilon_1=1$, %
for a field with a gravitation of like charge particles $\varepsilon_1=-1$. Let us note that EMT of a scalar field in form of(\ref{Tik_s})
is obtained from the Lagrangian \cite{YuNewScalar3}:
\begin{equation}\label{Ls}
L_s= \frac{\epsilon_1}{8\pi}(\Phi_i\Phi^{,i}-\epsilon_2 m^2_s\Phi^2).
\end{equation}
A requirement of total EMT conservation with an account of transport equations of particles' EMT leads to non-trivial relations such as equation of a scalar field with a source:
\begin{equation}\label{Eq_S}
\square\Phi+\epsilon_2m^2_s\Phi=-4\pi\epsilon_1\sigma
\end{equation}
and transport equation (\ref{GrindEQ__57_}).
For macroscopic set of equations' closure it is required to add definitions of particles' EMT and scalar charge density.

\section{A Thermodynamic Equilibrium And A Cosmological Model}
\subsection{Equilibrium Values Of Macroscopic Scalars}
 If colliding particles' free motion time $\tau _{eff} $ is sig\-ni\-fi\-cantly smaller than a typical time scale of a statistical system's evolution then the local thermo\-dy\-namic equilibrium (LTE) is maintained in the statistical system.
 At LTE conditions particle dis\-t\-ri\-bu\-tion functions take locally equilibrium form [5]:
\begin{equation} \label{GrindEQ__63_} f_{a}^{0} (x,P)=\frac{1}{e^{-\gamma _{a} +(\xi ,P_{a} )} \pm 1}  \end{equation}
 where $\xi ^{i} (x)$ is a timelike vector
\begin{equation} \label{GrindEQ__64_} \xi ^{2} \equiv (\xi ,\xi )>0, \end{equation}
at conditions of LTE at that {\it a kinematic velocity of plasma} (see (37)):
\begin{equation} \label{GrindEQ__65_} v^{i} =\xi ^{i} /\xi  \end{equation}
coincides with a dynamic velocity and scalar
\begin{equation} \label{GrindEQ__66_} \theta (x)=\xi ^{-1}  \end{equation}
is a local temperature of plasma. Let us emphasize the circumstance that at LTE a local temperature $\theta $ and a macroscopic velocity $v^{i} $
are the same for all plasma components. Next, in \eqref{GrindEQ__63_} $\gamma _{a} =\gamma _{a} (x)$ is a reduced chemical
potential of plasma's $a$ - component, which is connected to the classical one
$\mu_{a} $ by the relation:
\[\gamma _{a} =\frac{\mu _{a} }{\theta } .\]
At conditions of LTE chemical potentials of a sta\-tis\-tical system in which reactions \eqref{GrindEQ__67_} proceed should
satisfy a set of algebraic equations of chemical equilibrium:
\begin{equation} \label{GrindEQ__67_} \sum _{A=1}^{m} \nu _{A} \gamma _{A} =%
\sum _{B=1}^{m'} \nu '_{B} \gamma _{B'}, \end{equation}
 where all reaction channels with these particles participation should be accounted.
 Energy - mo\-mentum tensor of particles \eqref{GrindEQ__47_} with respect to a locally equilibrium distribution function
  \eqref{GrindEQ__63_} takes a structure of energy - momentum tensor of the per\-fect liquid:
\begin{equation} \label{GrindEQ__68_} T_{p}^{ik} =({\rm {\mathcal E}}_{pl} +{\rm {\mathcal P}}_{pl} )v^{i} v^{k} -{\rm {\mathcal P}}_{pl} g^{ik} , \end{equation}
where ${\rm {\mathcal E}}_{pl} $ and ${\rm {\mathcal P}}_{pl} $ are summary energy density and pressure of plasma.
For equilibrium scalar densities  %
\eqref{GrindEQ__46_}, \eqref{GrindEQ__47_} and \eqref{GrindEQ__48_} with respect to the equilibrium distribution \eqref{GrindEQ__63_} we find:

\begin{eqnarray} \label{GrindEQ__69_}
n_{a} =\frac{2S+1}{2\pi ^{2} } m_{*}^{3} \int _{0}^{\infty } \frac{{\rm sh}^{2} x{\rm ch}
xdx}{e^{-\gamma _{a} +\lambda _{*} {\rm ch} x} \pm 1} ;
\end{eqnarray}
\label{GrindEQ__69a_}
\begin{eqnarray}
{\rm {\mathcal E}}_{pl} =\sum _{a} \frac{2S+1}{2\pi ^{2} } m_{*}^{4}
\int _{0}^{\infty } \frac{{\rm sh}^{2} x{\rm ch}^{2} xdx}{e^{-\gamma _{a} +
\lambda _{*} {\rm ch} x} \pm 1}; \\
{\mathcal P}_{pl} =\sum _{a} \frac{2S+1}{6\pi ^{2} } m_{*}^{4} %
\int_{0}^{\infty }\frac{{\rm sh}^{4} xdx}{e^{-\gamma _{a} +\lambda _{*} {\rm ch} x} \pm 1}; \\
T_{p} =\sum _{a} \frac{2S+1}{2\pi ^{2} } m_{*}^{2} \int _{0}^{\infty }
\frac{{\rm sh}^{2} xdx}{e^{-\gamma _{a} +\lambda _{*} {\rm ch} x} \pm 1};\\
\sigma =\sum _{a} \frac{2S+1}{2\pi ^{2} } qm_{*}^{3}
\int _{0}^{\infty } \frac{{\rm sh}^{2} xdx}{e^{-\gamma _{a} +\lambda _{*} {\rm ch} z} \pm 1} ,
\end{eqnarray}
where $\lambda _{*} =m_{*} /\theta $.

Thus in LTE state macroscopic scalars of a statistical system are determined by functions: $\gamma_a$ ($a=\overline{1,N}$), $\lambda$, $\Phi$.
It is necessary to account also the equations of chemical equilibrium (\ref{GrindEQ__67_}),%
in con\-se\-qu\-ence of which scalar functions
should be determined by conservation laws of the fundamental charges %
$e_a$, which in turn are the consequences of transport equations (\ref{GrindEQ__54_}) %
with substituted $\Psi_a=e_a$ (see e.g. \cite{YuNewScalar2}):
\begin{equation}\label{ji}
\nabla_iJ^i=\nabla_i \sum\limits_a e_an_a u^i=0.
\end{equation}
At the same time it is necessary to bear in mind that at LTE conditions chemical potentials
 of massless particles should be equal to zero while chemical potentials of antiparticles are equal by modulus
 and opposite by sign to chemical potentials of identical particles:
\begin{equation}\label{gamma}
\bar{\gamma}_a=-\gamma_a.
\end{equation}

\subsection{Macroscopic scalars for degenerate single-component Fermi system}
At complete degeneracy conditions:
\begin{equation}\label{1}
\theta\to 0.
\end{equation}
locally equilibrium fermions' distribution function takes a form of a step function \cite{Ignatev4}:
\begin{equation}\label{2}
f^0(x,P)=\chi_+(\mu-\sqrt{m_*^2+p^2}),
\end{equation}
where $\chi_+(z)$ is a step function (Heaviside function).

Therefore an integration of macroscopic
densities is representable in elementary functions \cite{Ignatev4}:
\begin{equation}\label{3}
n=\frac{1}{\pi^2}p_F^3;
\end{equation}
\begin{equation}\label{3a}{\displaystyle
\begin{array}{l}
{\rm {\mathcal E}}_{pl} = {\displaystyle\frac{m_*^4}{8\pi^2}}
\bigl[\psi\sqrt{1+\psi^2}(1+2\psi^2)\\[8pt]
-\ln(\psi+\sqrt{1+\psi^2})\bigr];
\end{array}}
\end{equation}
\begin{equation}\label{3b}{\displaystyle
\begin{array}{l}
{\mathcal P}_{pl} ={\displaystyle\frac{m_*^4}{24\pi^2}}
\bigl[\psi\sqrt{1+\psi^2}(2\psi^2-3) \\ [8pt]
+3\ln(\psi+\sqrt{1+\psi^2})\bigr];
\end{array}}
\end{equation}
\begin{equation}\label{3c}{\displaystyle
\begin{array}{l}
\sigma={\displaystyle\frac{q\cdot m_*^3}{2\pi^2}}\left[\psi\sqrt{1+\psi^2}-
\ln(\psi+\sqrt{1+\psi^2})\right];
\end{array}}
\end{equation}
where
\begin{equation}\label{psi}\psi=p_F/m_*
\end{equation}
is a ratio of the Fermi momentum to the effective mass\footnote{The
specifications see in \cite{Ignatev4}.}.

\subsection{The Cosmological Model}
Let us consider the formulated earlier self-consistent mathematical model with regard to cosmological situation for the space-flat Friedman model:
\begin{equation}\label{Freedman}
ds^2=dt^2-a^2(t)(dx^2+dy^2+dz^2).
\end{equation}
Therefore all macroscopic scalars (\ref{3a}) -- (\ref{3c}) are dependent on time through an explicit dependency on two functions $a(t)$ and
 $\Phi(t)$. %
In this case $v^i=\delta^i_4$ and the fermion number conservation law takes form:
\begin{equation}\label{pfa}
ap_F={\rm Const}.
\end{equation}
This is the way the corresponding cosmological problem had been being solved earlier. However let us notice that the scalar charge conservation law in contrast to the electrical one does not follow from anywhere.
Therefore as opposed to previous articles we will not account it further.
Instead, it is necessary to use a plasma EMT transport law (\ref{GrindEQ__57_}), which for a cosmological situation takes form:
\begin{equation}\label{T,k(p)}
\dot{\rm {\mathcal E}}_{pl}+3\frac{\dot{a}}{a}({\rm {\mathcal E}}_{pl}+{\rm {\mathcal P}}_{pl})=\sigma\dot{\Phi}.
\end{equation}
Next, EMT of a scalar field also takes form of the perfect isotropic fluid's energy-momentum tensor:
\begin{equation} \label{MET_s}
T_{s}^{ik} =({\rm {\mathcal E}}_s +{\rm {\mathcal P}}_{s} )v^{i} v^{k} -{\rm {\mathcal P}}_s g^{ik} ,
\end{equation}
having:
\begin{eqnarray}\label{Es}
{\rm {\mathcal E}}_s=\frac{\epsilon_1}{8\pi}(\dot\Phi^2+\varepsilon_2 m_s^2\Phi^2);\\
\label{Ps} {\rm {\mathcal P}}_{s}=\frac{\epsilon_1}{8\pi}(\dot\Phi^2-
\varepsilon_2 m_s^2\Phi^2),
\end{eqnarray}
so that:
\begin{equation}\label{e+p}
{\rm {\mathcal E}}_s+{\rm {\mathcal P}}_{s}=\frac{\epsilon_1}{4\pi}\dot{\Phi}^2.
\end{equation}
The scalar field equation takes form:
\begin{equation}\label{Eq_S_t}
\frac{1}{a^3}\frac{d}{dt}a^3\Phi+\epsilon_2 m^2_s\Phi= -4\pi\epsilon_1\sigma(t).
\end{equation}
The last non-trivial Einstein equation must be added to these equations%
\begin{equation}\label{Einstein_a}
3\frac{\dot{a}^2}{a^2}=8\pi{\rm {\mathcal E}},
\end{equation}
where ${\rm {\mathcal E}}$ is a summary energy density of a Fermi system and a scalar field. This set of equations describes the closed mathematical model of cosmo\-lo\-gical evolution of the completely degenerate Fermi system with a scalar interaction.

In the next article we will carry out a numerical integration of the formulated here set of equations.

\end{document}